%% file: acl_latex.tex
\pdfoutput=1

\documentclass[11pt]{article}
\usepackage{enumitem}
\usepackage[preprint]{acl}

\usepackage{times}
\usepackage{latexsym}
\usepackage{graphicx}

\usepackage[T1]{fontenc}

\usepackage[utf8]{inputenc}

\usepackage{microtype}

\usepackage{inconsolata}

\usepackage{graphicx}

\title{Enhancing Leakage Attacks on Searchable Symmetric Encryption Using LLM-Based Synthetic Data Generation}

\author{
 \textbf{Joshua Chiu\textsuperscript{1}},
 \textbf{Partha Protim Paul\textsuperscript{1}},
 \textbf{Zahin Wahab\textsuperscript{1}}\\
 \textsuperscript{1}The University of British Columbia\\
}
\usepackage{enumitem}
\usepackage{booktabs}
\usepackage{xcolor}
\usepackage{listings}

\usepackage{xparse}

\NewDocumentCommand{\codeword}{v}{%
\texttt{\textcolor{blue}{#1}}%
}

\begin{document}
\maketitle
\begin{abstract}
\input{latex/abstract}
\end{abstract}

\section{Introduction}

\input{latex/introduction}

\section{Background}\label{sec:background}

Searchable symmetric encryption (SSE) is a specialized cryptographic scheme designed to enable keyword searches over encrypted data without exposing the underlying plaintext. This technology is used for applications such as cloud storage, where sensitive data is stored on potentially untrusted third-party cloud servers. SSE provides a balance between confidentiality and usability by allowing search operations while preserving data privacy \cite{curtmola2006sse}. In practice, these systems often leak ancillary information—such as search patterns and document access frequencies—that adversaries can exploit to reconstruct sensitive details from encrypted datasets \cite{cash2013}.

In a typical SSE scheme, the documents are encrypted and an encrypted index is constructed. This index is typically designed as an inverted index structure that maps keywords to the documents containing them. By encrypting both the documents and the index, the scheme ensures that an adversary on the server, or the server itself, cannot learn sensitive information about the stored data. However, some minimal leakage, such as the structure of the index or the frequency of keyword occurrences, may still occur. This data forms the foundation of most SSE attacks. 

The matching of the search tokens (or keywords) to the documents can be statistically attacked if there is an understanding of the domain of the database. In a database of encrypted medical records, it may be the case that the keyword "cancer" appears in 40\% of the documents, related to the number of cancer patients in the country. With occurrence information, attacks can be run to recover the search tokens (query recovery), and from that, the database can be rebuilt. 

In most experiments revolving around this scheme, the attacker is unrealistically assumed to have half of the datasets as leaked documents, with the other half serving as the "real" encrypted database. In a realistic sense, this means that half of the documents are decrypted and leaked in the scheme. In our experiment, we are trying to break this assumption with less, if not none, documents given to the attacker to perform the attack. We augment the fewer document count with generation from a large language model.

\section{Gap Analysis}\label{sec:gap_analysis}
Despite significant advancements in searchable symmetric encryption (SSE), these systems remain vulnerable to statistical leakage. Research has shown that adversaries can exploit access and search pattern leaks to reconstruct sensitive information. However, many existing studies operate under the assumption that attackers already possess a substantial portion—often 50\%—of the encrypted dataset. In practice, this assumption is unrealistic, as real-world attackers typically have access to only a small subset of the data, if any at all.

Large language models (LLMs) have demonstrated remarkable capabilities in generating synthetic data, particularly in privacy-preserving applications. Yet, little attention has been given to their potential role in enhancing adversarial attacks on SSE. This work aims to fill this gap by systematically evaluating how adversaries can leverage LLMs to generate auxiliary data, reducing their reliance on leaked real documents while maintaining or even improving attack effectiveness.

\section{Methodology}\label{sec:methodology}
\input{latex/methodology}

\section{Evaluation}\label{sec:eval}
\input{latex/result_Analysis}

\section{Threats to Validity}\label{sec:threats}
There are several threats to validity that should be considered in our study:

\begin{itemize}[leftmargin=*]
    \renewcommand{\labelitemi}{-}
    \item \textit{Model Selection Bias:} Our experiments were conducted solely using GPT-4o and GPT-4o mini. The behavior and performance of other large language models (LLMs) may differ, potentially affecting the generalizability of our findings across different architectures or training paradigms.

\item \textit{Data Contamination:} The Enron dataset, used as part of our evaluation, was publicly released in 2004 and is widely used in both academic and industry settings. It is possible that this dataset were included fully or partially in the pretraining corpus of the LLMs used. This could introduce unintended bias or leakage, especially if the model has memorized parts of the dataset. \end{itemize}
\section{Self Evaluation}\label{sec:self_eval}
\input{latex/self_evaluation}

\section{Conclusion and Future Work}\label{sec:conclusion}
\input{latex/conclusion}
\section*{Availability}
Data supporting this study's findings are available at \url{https://github.com/Joshua-Chiu/CPSC532V-LLM-SSE-Project}.

\bibliography{ref} 




\end{document}

%% file: latex/abstract.tex
Searchable Symmetric Encryption (SSE) enables efficient search capabilities over encrypted data, allowing users to maintain privacy while utilizing cloud storage. However, SSE schemes are vulnerable to leakage attacks that exploit access patterns, search frequency, and volume information. Existing studies frequently assume that adversaries possess a substantial fraction of the encrypted dataset to mount effective inference attacks, implying there is a database leakage of such documents, thus, an assumption that may not hold in real-world scenarios. In this work, we investigate the feasibility of enhancing leakage attacks under a more realistic threat model in which adversaries have access to minimal leaked data. We propose a novel approach that leverages large language models (LLMs), specifically GPT-4 variants, to generate synthetic documents that statistically and semantically resemble the real-world dataset of Enron emails. Using the email corpus as a case study, we evaluate the effectiveness of synthetic data generated via random sampling and hierarchical clustering methods on the performance of the SAP (Search Access Pattern) keyword inference attack restricted to token volumes only. Our results demonstrate that, while the choice of LLM has limited effect, increasing dataset size and employing clustering-based generation significantly improve attack accuracy, achieving comparable performance to attacks using larger amounts of real data. We highlight the growing relevance of LLMs in adversarial contexts.

%% file: latex/introduction.tex
Searchable Encryption (SE) has become a critical area of research due to the increasing reliance on cloud services for storing sensitive data such as healthcare records and personal emails. While encryption ensures confidentiality, it also limits the ability to retrieve specific information efficiently. To address this, Searchable Symmetric Encryption (SSE) \cite{curtmola2006sse} allows users to search encrypted data without revealing its contents to the cloud provider.

A typical SSE setup involves a client (Alice), who encrypts and uploads a set of sensitive documents to an untrusted server (Bob). Since Alice does not trust Bob with plaintext data, the documents are stored in encrypted form along with an encrypted index, which enables keyword-based searches. To retrieve relevant documents, Alice generates a trapdoor (an encrypted query) using a secret function and sends it to Bob. The server can then use this trapdoor to search the encrypted index and return the matching encrypted documents, without knowing the actual keywords. This ensures that Bob, despite being an honest-but-curious adversary, cannot learn the content of Alice’s documents or queries.

However, SSE schemes are vulnerable to leakage attacks \cite{cash2015}. Even though Bob cannot see the plaintext data, he observes query patterns, search frequency, and access patterns (i.e., which documents are retrieved in response to each query). An attacker can exploit these leakages in several ways:
\begin{itemize}[leftmargin=*]
    \renewcommand{\labelitemi}{-}
    \item \textit{Search Pattern Leakage:} If the same trapdoor is sent multiple times, Bob can infer that Alice is searching for the same keyword.
    \item \textit{Access Pattern Leakage:} By observing which documents are accessed for different queries, Bob can determine relationships between keywords and infer document contents \cite{islam2012access}.
    \item \textit{Volume Leakage:} If a query retrieves a large number of documents, Bob can guess that the keyword is common in the dataset.

\end{itemize}

These leakages allow an adversary to mount frequency analysis, statistical inference, and known-term attacks, significantly weakening the privacy guarantees of SSE.

Searchable symmetric encryption enables efficient search over encrypted data without revealing the plaintext, allowing for the storage of private data on untrusted cloud hosts. However, practical implementations of SSE are known to leak some statistical information (search patterns or access patterns) that adversaries can exploit. While SSE systems are designed to protect sensitive data, they are not immune to attacks through other aspects. Researchers have demonstrated various attack strategies that uses this leakage to recover sensitive information, and allows for the start of database reconstruction \cite{islam2012access}. 

Simultaneously, advances in natural language processing (NLP) have led to the development of powerful language models capable of generating realistic synthetic data \cite{vaswani2017attention,brown2020language}. These models can produce large volumes of high-quality text that mimics complex real-world dialogue patterns and document structures \cite{bommasani2021opportunities}. The intersection of these two domains allows for the potential of using synthetic data to simulate realistic environments, analyze vulnerabilities, and ultimately enhancing the sophistication of SSE attacks through the reduction of the assumptions of the attacker.
This paper explores leakage attacks on SSE using LLM-based synthetic data generation, demonstrating how advanced language models can be leveraged to simulate and exploit these vulnerabilities effectively.

 In this study, we utilized \codeword{gpt4o-mini} and \codeword{gpt4o} language models to generate enhanced datasets to aid the leakage attack. We compared three approaches: a baseline using only a limited original attacker dataset, random document augmentation, and clustering-based augmentation. Our results demonstrate that increasing the attacker's dataset size, particularly through clustering, significantly improves attack accuracy compared to the baseline. While switching from \codeword{gpt4o-mini} to \codeword{gpt4o} provided minimal gains, clustered enhancement consistently outperformed random augmentation by better capturing the statistical properties of the client's data. 

In summary, we make the following contributions:
 
\begin{itemize}[leftmargin=*]
    \renewcommand{\labelitemi}{-}
    \item We conduct leakage attacks on SSE using LLM-based synthetic data, thus reducing reliance on leaked real documents.
\item We show that LLM-generated synthetic data improves attack effectiveness.

\end{itemize}

The rest of the paper is organized as follows: Section \ref{sec:background} presents the background and related works, while Section \ref{sec:gap_analysis} points out the gap in current research. Then Section \ref{sec:methodology} details the methodology for the experimental setup. Section \ref{sec:eval} discusses the attack model, evaluation metrics and the baseline. Section \ref{sec:results} and provides a comprehensive analysis of experimental results. Section \ref{sec:threats} discusses potential threats to validity. Finally, Section \ref{sec:self_eval} sheds light into our self evaluation of the project, and Section \ref{sec:conclusion} concludes the paper and points out the future direction for this research.

%% file: latex/methodology.tex
\paragraph{Dataset.} We used the Enron email dataset \cite{shetty2004enron}, a well-known collection of real-world emails originally made public during the Enron Corporation investigation. This dataset consists of around 500,000 email communications exchanged between Enron employees, providing a valuable resource for studying natural language patterns, workplace communication, and privacy-related concerns. Inline with previous research on this topic, we will filter and use all the emails in the \codeword{_sent_items} \cite{oya2020, oya2022, jigsaw2024, islam2012access}.

\paragraph{Dataset Sanitization.} To maintain data consistency, we preprocessed the Enron dataset to remove metadata and header information such as sender, recipient, timestamp, and routing data, retaining only the email content to ensure a standardized plain text textual format.

\paragraph{Data Selection.} We divide the sanitized dataset into two equal parts:

\begin{itemize}[leftmargin=*]
   \renewcommand{\labelitemi}{-}
 \item \textit{Client Data:} Represents encrypted user data that is assumed to be private and secure.
 \item \textit{Attacker Data:} Represents data that has been leaked and potentially accessible to an adversary.

\end{itemize}
However, we assume that the attacker does not have access to the full leaked data. Instead, the attacker possesses only a subset of this leaked information. Using this limited knowledge, the attacker attempts to generate synthetic data to approximate the remainder of the attacker data portion. This setup models a realistic threat scenario where the adversary's knowledge is incomplete, and they rely on data generation or inference techniques to expand their dataset.

We then randomly select a subset of emails from the attacker set of the sanitized Enron dataset for analysis and synthetic data generation. We ran this experiment multiple times with different random seeds to strengthen the findings.

\paragraph{LLM under Test.} We used \codeword{gpt-4o} and \codeword{gpt-4o-mini} \cite{openai2023gpt4} as the language model for synthetic data generation that mimics the statistical properties and linguistic characteristics of the real Enron dataset. These models fit our criteria as it has been trained on diverse datasets, allowing it to learn linguistic patterns, structures, and contextual relationships.

\paragraph{Synthetic Attacker Data Generation.} For enhancing the attacker's data, we used a fixed proportion: say 20\% real data (from the initial 50\% leaked dataset) combined with 80\% synthetic data generated using \codeword{GPT}. We simulated a scenario starting with an initial `seed dataset', representing 20\% of a larger, original data distribution. The goal was to assess how effectively properties (keywords) of the original distribution could be inferred by analyzing this seed set after augmenting it. The augmentation process involved using only the real seed data (the initial 20\%) to prompt \codeword{GPT}, generating synthetic data to form a final analysis set composed of a fixed mix: 20\% real seed data and 80\% synthetic data. We investigated the impact of scale on inference effectiveness by varying the size of this initial seed dataset across experiments (e.g., starting with 50 seed emails, up to 2000 seed emails), which implies corresponding variations in the size of the conceptual original dataset, always assuming a 20\% initial leak.

To generate synthetic data, we used the following prompt mentioned in the table \ref{tab:prompt} with \codeword{GPT}:

\begin{table}[ht]
\centering
\caption{Prompt used for Synthetic Data Generation.}
\label{tab:prompt}
\resizebox{\columnwidth}{!}{%
\begin{tabular}{|c|}

\hline
\texttt{Generate \{message\_per\_cluster\} new texts similar to the following examples:} \\
\texttt{Example 1: \{example\_1\}} \\
\texttt{Example 2: \{example\_2\}} \\
\texttt{Example 3: \{example\_3\}} \\
\texttt{Return only the response as a separate entry and put exactly '\#\#\#' between them so I can parse them.} \\
\texttt{Do not generate any extra messages.} \\
\hline
\end{tabular}
}

\end{table}

Here, \codeword{{message_per_cluster}} specifies the number of texts to generate per cluster, and each \codeword{{example_i}} is a representative sample from the cluster.

\paragraph{Hierarchical Clustering.} 
To structure the synthetic data generation process, we first applied hierarchical clustering\cite{simeone2023incrementalhierarchicaltextclustering} to the selected real data before feeding it into \codeword{GPT}. This step allowed us to identify meaningful groupings within the dataset, ensuring that the generated data preserved its inherent diversity and structural patterns. From each identified cluster, we randomly selected a small number of real emails (say, three to five). These selected emails were then used as input for \codeword{GPT}, which generated between five and ten new emails per cluster (the exact number determined based on the overall quantity of synthetic data required). By leveraging hierarchical clustering, this approach maintained the natural distribution of the original dataset while enhancing the adversary’s ability to infer and reconstruct the complete database from limited leaked data.

\paragraph{Random Selection.}
In contrast to the hierarchical clustering approach, the random selection method did not rely on any form of data grouping. Instead, we selected real data points randomly from the available dataset without considering their structural relationships. We used the randomly chosen samples as prompts for \codeword{GPT}. \codeword{GPT} then generated between five and ten similar data points for each selected sample. This process continued until we reached the target dataset size. This method served as an unstructured baseline to evaluate the effectiveness of synthetic data in adversarial attacks. It allowed us to compare the impact of structured versus random selection on data reconstruction and inference accuracy.

\paragraph{Stemming.}

In constructing the search index, most Searchable Symmetric Encryption (SSE) schemes apply stemming techniques to ensure that query functionality aligns with user expectations and increasing usability—for instance, a query for "run" should retrieve documents containing variations such as "running". As such, and in accordance with established methodology, we preprocess each document by converting it into a set of keyword stems \cite{oya2020, islam2012access}.

In particular, we perform the following steps:
\begin{enumerate}[topsep=0pt,itemsep=-1ex,partopsep=1ex,parsep=1ex]
    \item a document is split into list of words
    \item all words are converted to lowercase
    \item words containing non-alphabetic characters are discarded
    \item words that are English stopwords are discarded
    \item words whose lengths fall outside the range of 3 to 20 characters are discarded
    \item words are processed using the Porter stemming algorithm
\end{enumerate}
After this process, a document that is originally \codeword{``The quick brown fox jumps over the lazy dog"} becomes \codeword{{`jump', `brown', `fox', `lazi', `dog', `quick'}}. 
These resulting normalized set of stemmed keywords representing each document are used as the basis for evaluation in our study.

%% file: latex/result_Analysis.tex
\paragraph{Attack Model.} 
As this experiment evaluates the ability of language models to generate synthetic data that can be used to perform attacks, we will evaluate the effectiveness through an access pattern-based query recovery attack that exploits keyword frequency information. We chose the SAP attack as it primarily exploits the statistical properties of keyword frequencies \cite{oya2020}. These scheme focus on using the access pattern, which
refers to the set of documents that match the client’s queries. This allows us to see the effectiveness of assuming the attacker having less data, but a strong knowledge of the domain and employing language model tools to enhance the auxiliary data used for the attack.

Consider a client-server model in which the client possesses a private database and seeks to outsource its storage to a server in order to reduce local storage overhead. Instead of downloading the whole database everytime, the client employs a privacy-preserving Searchable Symmetric Encryption (SSE) scheme which retains the capability to perform point queries. The scheme operates as follows. Initially, the client encrypts the database using a symmetric encryption algorithm and transmits the encrypted data to the server, accompanied by a query index. Subsequently, when the client wishes to search for a specific keyword, it generates a query token and sends it to the server. The server then evaluates this token against the query index to retrieve the addresses of documents that correspond to the search criterion. Notably, although the server does not compromise the security guarantees of the scheme, it may still infer the number of documents that match each query token \cite{oya2020}. In large sample, the volumes of each query token should regress to the Zipfian distribution of the english language.

\begin{figure}[h]
    \centering
    \includegraphics[width=0.48\textwidth]{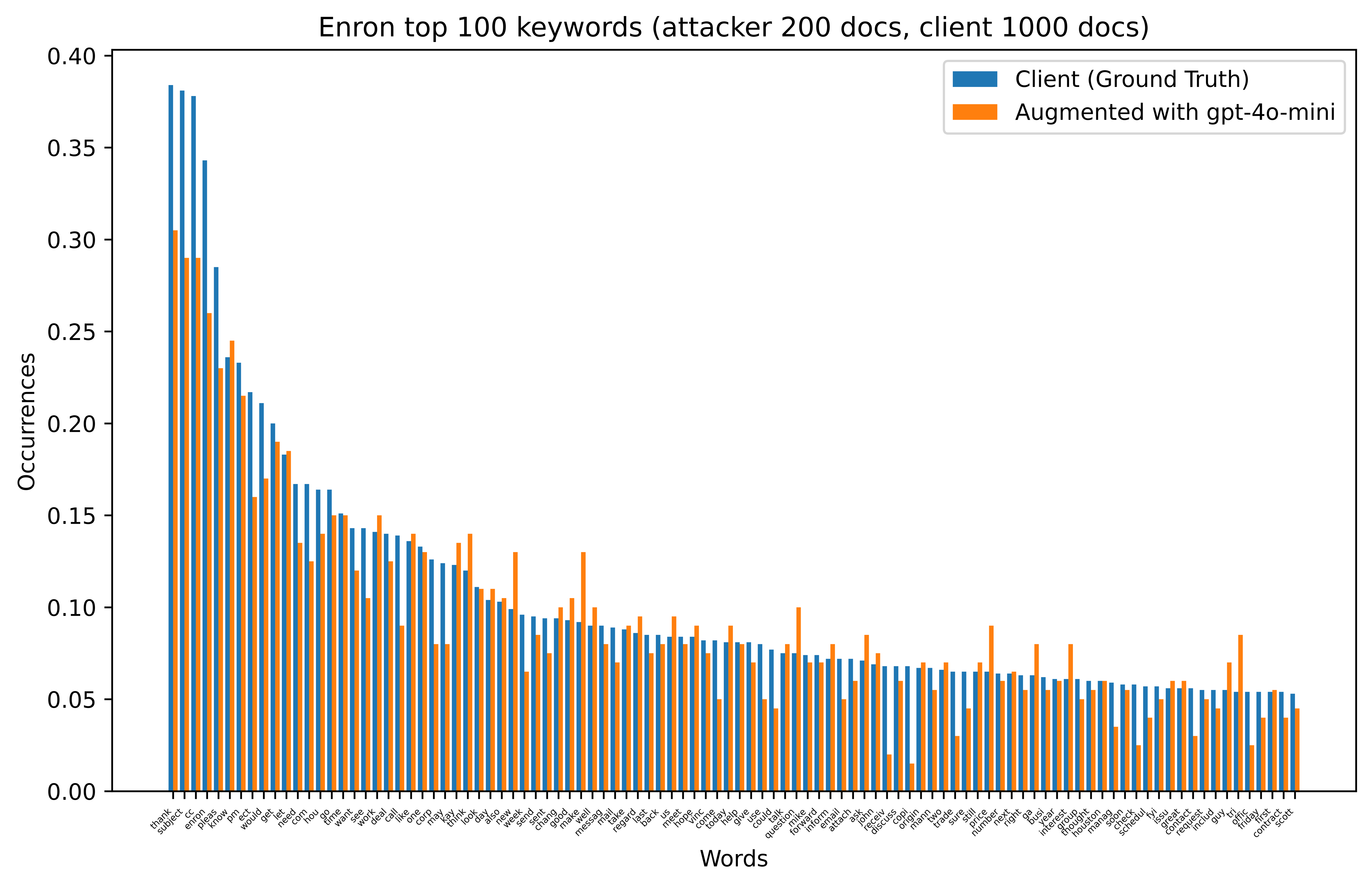}
    \caption{Bar chart showing the client's word frequencies sorted in descending order. For each word, the corresponding frequency from the attacker is shown in the same position.}
    \label{fig:zipfian distribution}
\end{figure}

Figure \ref{fig:zipfian distribution} shows a typical word frequency distribution where the client's word frequencies, when sorted in descending order, approximate a Zipfian distribution, and compare it to the attacker's corresponding frequency of the keywords.

\paragraph{Jaccardian Similarity.} The primary metric will be the Jaccardian similarity of the two matchings of the top 100 recovered keywords, where matchings are simply the sorted keywords occurrence. This allows for an accurate comparison even if the document count is different (i.e. when attacker and client data are not of same size). This metric is a simple and basic yet effective metric in identifying whether any improvements in attack effectiveness are made. 

The Jaccardian similarity is a measure of the overlap between two sets \( A \) and \( B \). It is defined as:

\[
J(A, B) = \frac{|A \cap B|}{|A \cup B|}
\]
where $|A \cap B|$ is the number of elements common to both sets, and $|A \cup B|$
 is the total number of distinct elements in either set. A value of 1 indicates perfect similarity, while a value of 0 means the sets have no elements in common. Say if $A$ is the attacker data and $B$ is the client data, then the higher the value of Jaccardian similarity is, the more successful the attacker is in inferring client's data.

\paragraph{Baselines.} Our baseline comparison will be against using the real data to perform the attack where we execute the access pattern-based query recovery attack using all of the selected subsets of the attacker's dataset as the source of keyword frequency information compared to the client dataset. Note that the attack accuracy of this is not perfect and is around 20-50\% \cite{cash2015, oya2020}. Our goal is to compare against the accuracy if the attacker had no access to a language model and used just the original real documents.

\section{Results}\label{sec:results}

To assess the effectiveness of our enhanced dataset, we evaluate its performance in comparison to the original, smaller dataset by executing the Search Access Pattern (SAP) attack. The SAP attack is performed using only volume information, specifically the number of documents matched by each query token. It is important to note that, in our experimental setup, the attacker does not have access to the client’s data. The attacker only possesses a limited amount of real data and generates synthetic data using a large language model (LLM) to match the client’s data. The attack’s success is measured using Jaccardian similarity, which compares the attacker’s generated data to the unseen client data to assess how accurately the attacker can approximate the client’s keyword distribution. We compare the performance of three different approaches based on if and how synthetic data is generated:
\begin{enumerate}[leftmargin=*]
    \item
\codeword{None:} This approach evaluates attack accuracy using only real attacker documents, measuring the Jaccardian similarity between the real attacker data and the client data.
\item
\codeword{Random Enhanced:} In this approach, the attack accuracy is evaluated using real attacker documents, supplemented by additional documents generated by a large language model (LLM) through random selection. The Jaccardian similarity is then calculated between the total attacker data (real + LLM-generated) and the client data.
\item
\codeword{Clustered Enhanced:} This approach evaluates attack accuracy in a similar way to the "Random Enhanced" approach, but the additional documents are generated using an LLM through hierarchical clustering. The Jaccardian similarity of the total attacker data (real + LLM-generated) and the client data is then computed.

\end{enumerate}
\input{latex/result_table}

\subsection{Varying Ratio for Attacker's Real to Synthetic Data}

Initially, we experimented with different ratios of real-to-synthetic data using only the \codeword{gpt-4o-mini} model on a smaller dataset. We gradually varied the proportion of real and synthetic data to observe its effect on the performance of the SAP attack. As shown in Table~\ref{tab:overall-comparison}, the performance of the SAP attack improves as the amount of synthetic data generated by the LLM increases. The results are summarized as follows:
\begin{itemize}[leftmargin=*]
 \renewcommand{\labelitemi}{-}
\item For the \codeword{gpt-4o-mini} model with 100 real and 900 synthetic documents, the \codeword{None} approach achieves a Jaccardian similarity of 0.2430. By incorporating randomly generated LLM-data, the similarity slightly decreases to 0.2106 due to noise, but clustering-based augmentation improves the similarity to 0.2442.

\item With the increase in the amount of real attacker data, Jaccardian similarity improves, even without augmentation of synthetic data. For instance, with 400 real documents, \codeword{None} yields a similarity of 0.2656, compared to 0.2430 with 100 real documents. 

\item As the real-to-synthetic ratio approaches 1:1 (500 real, 500 synthetic), \codeword{Clustered Enhanced} reaches a similarity score of 0.2993, outperforming both \codeword{None} (0.2707) and \codeword{Random Enhanced} (0.2917).

\end{itemize}
These results indicate a positive correlation between the synthetic data ratio and the attack accuracy, particularly with the \codeword{Clustered Enhanced} approach, which consistently performs the best across all configurations. Moreover, increasing the real attacker data improves the performance of the \codeword{None} approach, but the improvement is smaller compared to the enhancements provided by LLM generated data (\codeword{Random} and \codeword{Clustered Enhanced}).

\subsection{Using 1:4 Ratio for Attacker's Real to Synthetic Data}

Due to cost constraints, we were unable to experiment with different dataset size across all ratios. So, we fix the attacker-to-synthetic data ratio at 1:4 and vary the dataset sizes for both the \codeword{gpt-4o-mini} and \codeword{gpt4o} models. The second part of Table~\ref{tab:overall-comparison} presents the performance of these models under this fixed ratio, comparing the impact of model choice and dataset size on the effectiveness of keyword inference attacks, as measured by the Jaccard similarity score. In this setup, we assume the attacker has only 20\% real data and generates the remaining 80\% using the LLM. Specifically, we assessed the performance of different language models in generating attacker data and measured their influence on attack accuracy. Each experiment was repeated five times, and the results were averaged to account for any stochastic variation in document generation by the language models.

\begin{itemize}[leftmargin=*]
    \renewcommand{\labelitemi}{-}
    
    \item \textit{Model Comparison:} When attacker and client dataset sizes are identical, the resulting Jaccard Similarity scores are nearly identical across all enhancement strategies. This is particularly true under the "No Enhancement" setting, where the attacks operate on unmodified real data. Despite enhancements like randomized document generation or clustering-based augmentation, performance gains are similar between models. This suggests that switching to a more powerful model (e.g., \codeword{gpt4o}) results in only marginal improvements in keyword recovery, and these differences are statistically insignificant in most settings (more details in Section \ref{sec:stat}).

    \item \textit{Impact of Dataset Size:} In contrast to model choice, increasing the dataset size has a substantially more significant effect on attack accuracy. When both attacker and client datasets are scaled by an order of magnitude (from 200/1000 to 2000/10000), we observe a notable increase in Jaccard Similarity across all enhancement strategies. For instance, without any enhancement, the similarity score improves from 0.2537 to 0.3822. When clustering is applied to augment the dataset using \codeword{gpt-4o-mini}, this score increases further to 0.3963. These results indicate that larger datasets provide the attacker with a better approximation of the client’s keyword distribution, significantly improving inference accuracy.

    \item \textit{Effect of Randomly Generated Documents:} Interestingly, we observe that randomly generated additional documents can sometimes degrade performance. For example, when the attacker size is 2000 and the client size is 10000, the performance is slightly worse than when no enhancement is applied. This is likely due to the introduction of topic-incoherent noise, which harms the model's ability to infer relevant keywords.

    \item \textit{Effect of Clustering-Based Augmentation:} On the other hand, document clustering, where generated content is grouped and expanded based on coherent topical structure, consistently enhances attack effectiveness. Although the improvement tends to show diminishing returns as dataset size increases, clustering remains the most reliable and impactful enhancement strategy. This approach allows for more accurate modeling of the semantic and statistical properties of the client dataset, leading to improved keyword recovery accuracy.

\end{itemize}

To summarize, the results demonstrate that increasing dataset size significantly improves keyword inference accuracy, with larger datasets providing better approximations of the client’s keyword distribution. While model choice has a minor impact, generating synthetic data via clustering (\codeword{Clustered Enhanced)} consistently outperforms others, showing the greatest effectiveness in augmenting data and improving attack performance. Random enhancement, though useful, sometimes introduces noise, but clustering remains the most reliable and impactful strategy for enhancing keyword recovery and attack efficiency.

\subsection{Statistical Analysis}
\label{sec:stat}

We applied different statistical tests based on the distribution of the data to analyze the impact of different enhancement techniques and models. 

\paragraph{Impact of Enhancement. }The \texttt{Wilcoxon signed-rank test} was used to compare \codeword{None} with both \codeword{Random Enhanced} and \codeword{Clustered Enhanced} because the data for these comparisons was not normally distributed (as confirmed by normality tests). The \texttt{Wilcoxon signed-rank test} is a non-parametric test that compares the ranks of paired differences and is ideal for non-normally distributed data.

On the other hand, the \texttt{paired t-test} was applied to compare \codeword{Random Enhanced} with \codeword{Clustered Enhanced} because the differences between these two approaches were normally distributed, as confirmed by normality tests. The \texttt{paired t-test} is a parametric test that compares the means of two related groups, assuming normality, and is more powerful in this case.

\begin{itemize}[leftmargin=*]
    \renewcommand{\labelitemi}{-}
    \item \textit{Wilcoxon signed-rank test:} The significant p-values (0.030 for \codeword{None} vs \codeword{Random Enhanced} and 0.00006 for \codeword{None} vs \codeword{Clustered Enhanced}) indicate that both enhancement strategies (LLM-generated data through random selection and hierarchical clustering) provide significant improvements in attack performance compared to using only real data.
    
    \item \textit{Paired t-test:} The \textit{p-value} of 0.00028 for \codeword{Random Enhanced} vs \codeword{Clustered Enhanced} confirms that the difference between these two enhancement strategies is statistically significant, with \codeword{Clustered Enhanced} outperforming \codeword{Random Enhanced}.
\end{itemize}

In summary, we used the \texttt{Wilcoxon test} for non-normally distributed data and the \texttt{paired t-test} for normally distributed data, and both tests showed significant improvements from the enhanced approaches over the baseline.
\paragraph{Impact of Model (gpt-4o vs gpt-4o-mini).}

For each combination of dataset size at the 1:4 real-to-synthetic data ratio and enhancement technique, we measured the p-value using the \texttt{Wilcoxon signed-rank test} between \codeword{gpt-4o} and \codeword{gpt-4o-mini}. Since the attacker’s real and synthetic data, as well as the enhancement technique, were kept constant across both models, the only difference was the model itself. 

The differences in Jaccardian similarity between \codeword{gpt-4o} and \codeword{gpt-4o-mini} were not normally distributed (as confirmed by normality test), so we proceeded with the \texttt{Wilcoxon signed-rank test}. The result from this test showed a \texttt{W-statistic} of 20.0 and a \texttt{p-value} of 0.4922, indicating that there is no statistically significant difference between \codeword{gpt-4o} and \codeword{gpt-4o-mini}. 

These findings suggest that using \codeword{gpt-4o} over \codeword{gpt-4o-mini} does not provide a significant improvement in generating synthetic data for enhancing the accuracy of the keyword reconstruction attack.

%% file: latex/result_table.tex
\begin{table*}[h]
\centering

\caption{Performance Comparison (Jaccard Similarity) of Different Approaches}
\label{tab:overall-comparison}
\resizebox{\textwidth}{!}{%
\begin{tabular}{@{}lccccccr@{}}
\toprule
\multicolumn{7}{c}{Attacker's Real to Synthetic Data (Varying Ratio)}\\
\toprule
\textbf{Model} & \textbf{Attacker Size (Real Data)} & \textbf{Attacker Size (Synthetic Data)} & \textbf{Client Size} & \textbf{None} & \textbf{Random Enhanced} & \textbf{Clustered Enhanced} \\
\midrule

gpt4o-mini &	100	& 900 & 1000	& 0.2430 &	0.2106 &	\textbf{0.2442} \\
gpt4o-mini&	200	& 800 &1000&	0.2537&	0.2594 &	\textbf{0.2623} \\
gpt4o-mini&	300& 700&	1000	&0.2591&	0.2201	&\textbf{0.2623} \\
gpt4o-mini&	400& 600&	1000	&0.2656	&0.2768	& \textbf{0.2771}\\
gpt4o-mini &	500	&500&1000&	0.2707	& 0.2917 &	\textbf{0.2993} \\
\midrule
\multicolumn{7}{c}{Attacker's Real to Synthetic Data (1:4)}\\
\toprule
\textbf{Model} & \textbf{Attacker Size (Real Data)} & \textbf{Attacker Size (Synthetic Data)} & \textbf{Client Size} & \textbf{None} & \textbf{Random Enhanced} & \textbf{Clustered Enhanced} \\
\midrule
gpt4o-mini & 50  & 200 & 250  & 0.2141 & 0.2619 & \textbf{0.2693} \\
gpt4o      & 50  & 200 &250  & 0.2141 & 0.2606 & \textbf{0.2717} \\

gpt4o-mini & 100 & 400 &  500  & 0.2444 & 0.2909 & \textbf{0.3015} \\
gpt4o      & 100   & 400 & 500  & 0.2444 & 0.2911 & \textbf{0.3046} \\

gpt4o-mini & 200 & 800  & 1000  & 0.2537 & 0.2901 & \textbf{0.3081} \\
gpt4o      & 200 & 800  & 1000  & 0.2537 & 0.2915 & \textbf{0.3088} \\

gpt4o-mini & 400  & 1600 & 2000  & 0.2732 & 0.3146 & \textbf{0.3169} \\
gpt4o      & 400  & 1600  & 2000 & 0.2732 & 0.3049 & \textbf{0.3156} \\

gpt4o-mini & 2000 & 8000 & 10000 & 0.3822 & 0.3790 & \textbf{0.3963} \\
gpt4o      & 2000 & 8000 & 10000 & 0.3822 & 0.3646 & \textbf{0.3959} \\
\bottomrule
\end{tabular}
}
\end{table*}

%% file: latex/self_evaluation.tex


Overall, the project successfully followed the methodology outlined in the initial proposal. We used the sanitized Enron email corpus and simulated a data leakage attack where the attacker has access to a smaller subset of the original dataset (the "seed/real data"). Using this seed data, we generated synthetic data with Large Language Models (\codeword{gpt-4o} and \codeword{gpt-4o-mini}), as planned. Due to cost constraints, we were unable to run experiments by varying the dataset size across all real to synthetic attacker data ratios, so we only varied dataset sizes for the 1:4 real-to-synthetic ratio to assess its impact.

One outcome we did not fully anticipate was the significant improvement in performance when synthetic data was augmented with hierarchical clustering. Initially, we hypothesized that synthetic data would aid the attacker, especially in sparse datasets, by improving keyword reconstruction. The results aligned with our expectations, but the performance boost was much greater than anticipated. This highlights the effectiveness of data augmentation, particularly when using clustering techniques to guide synthetic data generation.

%% file: latex/conclusion.tex
In summary, our findings demonstrate that while the choice of language model has negligible influence on attack accuracy, the size and semantic organization of the dataset in synthetic data generation play a critical role. Clustering-based augmentation, in particular, provides a consistent and substantial improvement across all evaluated scenarios.

Future work can explore the generalizability of our findings by evaluating a broader range of large language models beyond GPT-4o and GPT-4o mini, including models with different architectures and training data. This would help assess whether the observed behaviors are consistent across LLMs or specific to the ones we tested. Additionally, to reduce the risk of data contamination, future studies could use more recent or proprietary datasets that are unlikely to have been included in public model pretraining.